# System Identification of NN-based Model Reference Control of RUAV during Hover

Bhaskar Prasad Rimal[1], Idris E. Putro[2], Agus Budiyono[2],
Dugki Min[3] and Eunmi Choi[1]
*[1]Graduate School of Business IT, Kookmin University*
*Jeongneung-Dong, Seongbuk-Gu, Seoul, 136-702, Korea*
*[2]Department of Aerospace Information Engineering, Konkuk University*
*[3]School of Computer Science and Engineering, Konkuk University*
*Hwayang-dong, Gwangjin-gu, Seoul 13-701,*
*Korea*

## 1. Introduction

Unmanned aerial vehicles (UAVs) are becoming more and more popular in a wide field of applications nowadays. UAVs are used in number of military application for gathering information and military attacks. In the future will likely see unmanned aircraft employed, offensively, for bombing and ground attack. As a tool for research and rescue, UAVs can help find humans lost in the wilderness, trapped in collapsed buildings, or drift at sea. It is also used in civil application in fire station, police observation of crime disturbance and natural disaster prevention, where the human observer will be risky to fight the fire. There is wide variety of UAV shapes, sizes, configuration and characteristics. Therefore, there is a growing demand for UAV control systems, and many projects either commercial or academic destined to design a UAV autopilot were held recently. A lot of impressive results had already been achieved, and many UAVs, more or less autonomous, are used by various organizations.

An Artificial Neural Network (ANN) [3] is an information processing paradigm that is stimulated by the way biological nervous systems, such as the brain, process information. The key element of this paradigm is the novel structure of the information processing system. Basically, a neural network (NN) is composed of a set of nodes (Fig. 1). Each node is connected to the others via a set of links. Information is transmitted from the input to the output cells depending of the strength of the links. Usually, neural networks operate in two phases. The first phase is a learning phase where each of the nodes and links adjust their strength in order to match with the desired output. A learning algorithm is in charge of this process. When the learning phase is complete, the NN is ready to recognize the incoming information and to work as a pattern recognition system.

ANNs, like people, learn by example. An ANN is configured for a specific application, such as pattern recognition or data classification, through a learning process. Learning in biological systems involves adjustments to the synaptic connections that exist between the neurons.

In recent years, there is a wide momentum of ANNs in the control system arena, to design the UAVs. Any system in which input is not proportional to output is known as non-linear systems. The main advantages of ANNs are having the processing ability to model nonlinear systems. ANNs are very suitable for identification of non-linear dynamic systems. Multilayer Perceptron model have been used to model a large number of nonlinear plants. We can vary the number of hidden layers to minimize the mean square error. ANNs has been used to formulate a variety of control strategies [1] [2]. The NN approach is a good alternative for physical modeling techniques for nonlinear systems.

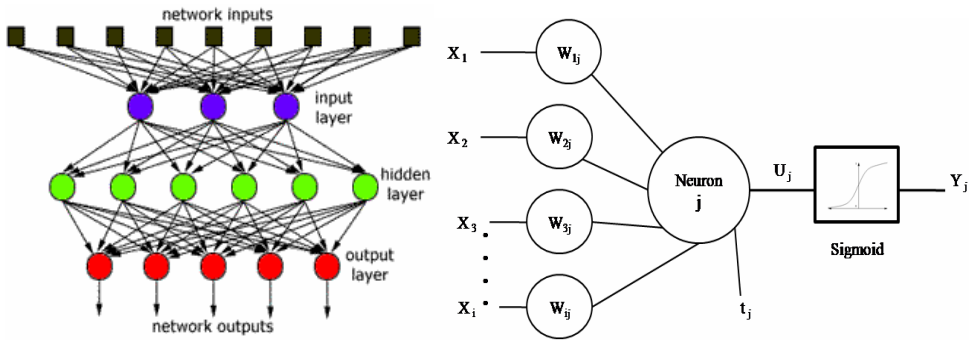

Fig. 1. General Neural Network Architecture

A fundamental difficulty of many non-linear control systems, which potentially could deliver better performance, is extremely difficult to theoretically predict the behavior of a system under all possible circumstances. In fact, even design envelope of a controller often remains largely uncertain. Therefore, it becomes a challenging task to verify and validate the designed controller under all possible flight conditions. A practical solution to this problem is extensive testing of the system. Possibly the most expensive design items are the control and navigation systems. Therefore, one of main questions that each system designer has to face is the selection of appropriate hardware for UAV system. Such hardware should satisfy the main requirements without contravening their boundaries in terms of quality and cost. In UAV design this kind of consideration is especially important due to the safety requirements expressed in airworthiness standards. Therefore question is how to find the optimal solution. Thus, simulation is necessary. Basically there are two type of simulation is needed while designing UAVs systems, they are Software-In-the-Loop (SIL) [5] simulation and Hardware-In-the-Loop (HIL) simulation [4].

To utilize the SIL configuration, the un-compiled software source code, which normally runs on the onboard computer, is compiled into the simulation tool itself, allowing this software to be tested on the simulation host computer. This allows the flight software to be tested without the need to tie-up the flight hardware, and was also used in selection of hardware.

HILS simulates (Fig. 2) a process such that input and output signals show the time-dependent values as real-time operating components. It is possible to test embedded system under real time with various test conditions. It provides the UAV developer to test many aspects of autopilot hardware, finding the real time problems, test the reliability, and many more.

The simulation can be done with the help of Matlab Simulink program environment. This program can be considered as a facility fully competent for this task. Simulink is the most popular tool, it was not only used for a SIL Simulation of the complete UAV system but also to create the simulation code of a HIL Simulator that runs in real time.

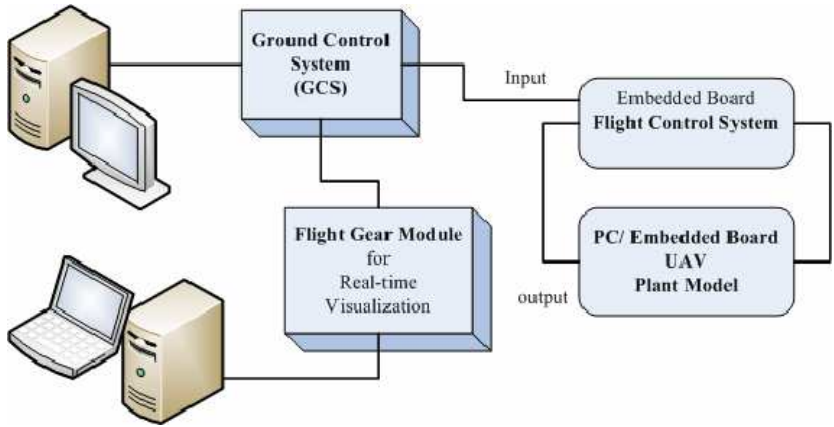

Fig. 2. UAV Architecture: Hardware-in-the-loop Simulation

The system identification is the first and crucial step for the design of the controller, simulation of the system and so on. Frequently it is necessary to analyze the flight data in the frequency domain to identify the UAV system. This paper demonstrates how ANN can be used for non linear identification and controller design. The simulation processes consists of designing a simple system, and simulates that system with the help of model reference control block in Matlab/Simulink [6].

The paper is organized as follows: Section 2 describes some related work. Section 3 deals with system identification and control on the basis of NNs. Details of design and control system with NNs approaches is describes in section 4. In section 5, simulations are performed on RUAVs system and finally, conclusions are drawn in section 6.

## 2. Related work

Robust control techniques are capable for adapting themselves for changing the dynamics which are necessary for autonomous flight. This kind of controller can be designed with the help of system identification.

There are lots of work had already done in UAV area in the context of ANNs. Mettler B. et al., [12] describe the process and result of the dynamic modeling of a model-scale unmanned helicopter using system identification. E. D. Beckmann et al., [13] explained the nonlinear modeling of a small-scale helicopter and the identification of its dynamic parameters using prediction error minimization methods. NN approaches have excellent performance than classical technique for modeling and identifying nonlinear dynamic systems [15] [16].

There is also numerous system identification techniques had been developed to model nonlinear systems. Some of them are Fuzzy identification [20] [27], state-space identification [21], frequency domain analysis [22], NN based identification [23] [26]. The exception is given by LPV identification [25] which is applicable for the entire flight envelope. The learning ability is the beauty of NN that has been utilized widely for system identification and control applications. Shim D. H. et al. [28] described time-domain system identification approaches to design the control system for RUAVs.

## 3. System identification and control

The main idea of system identification is often to get a model that can be used for controller design. System identification (SI) [7] provides the idea of making mathematical models of dynamics systems, starting from experimental data, measurements, and observations.

It is widely used for applications ranging from control system design and signal processing to time series analysis. The system identification is used to verify and test the control system parameters that are associated with the six-degree-of-freedom system using the test flight data. The simulation results and the statistical error analysis are provided for both the cases. Fig. 5 shows the flow of control system design with the system identification model. Basically System identification is the experimental approach to process modeling and it includes the following five steps as shown in Fig. 3

The system considered as a black box (Fig. 4) which receives some inputs that lead to some output. The concern here is: what kind of parameters for a particular black box can correlate the observed inputs and outputs?

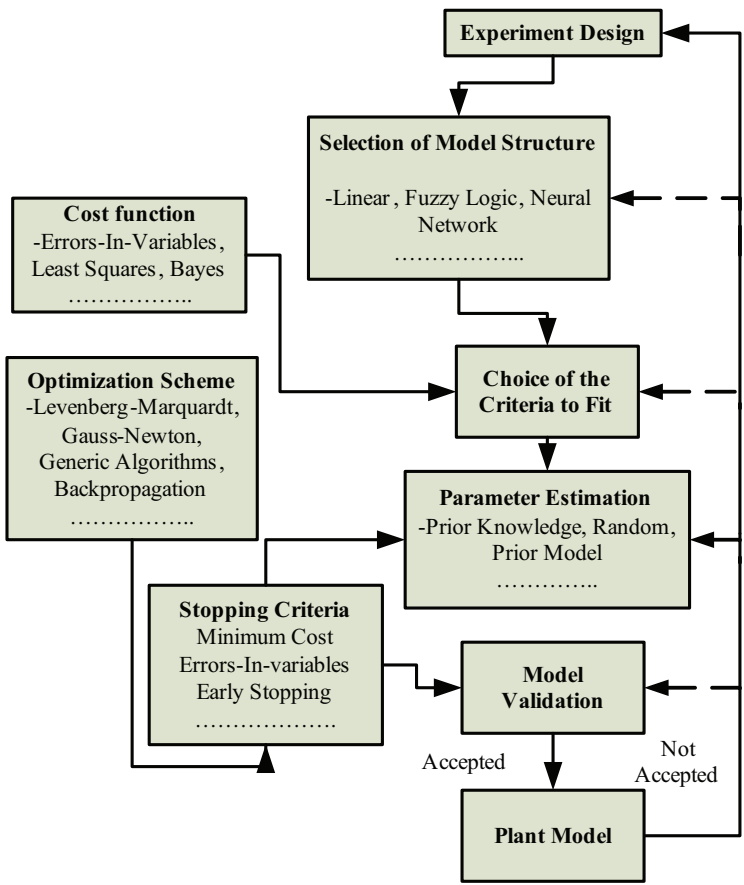

Fig. 4. System Identification Modeling Procedure

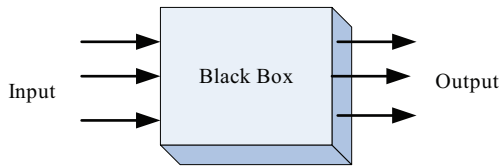

Fig. 5. Black box view of System Identification

Can these parameters help designer to predict the outputs for a new given set of inputs? This is the basic problem of system identification. Neural networks have been applied successfully in the identification and control of dynamic systems. Popular neural network architectures for prediction and control that have been implemented in the NN Toolbox™ software are:

- Model Predictive Control
- NARMA-L2 (Feedback Linearization) Control
- Model Reference Control

There are typically two steps involved when using neural networks for control systems:

- System identification
- Control design

In the system identification stage, we develop a neural network model of the plant that we want to control. The flow of control system design with system identification is shown in Fig 5. In the control design stage, we use the neural network plant model to design (or train) the controller using the propagation of the controlling error through the NN model. Training produces the optimal connection weights for the networks by minimizing the errors between NN output and the plant output over the entire set of samples. Among many network training algorithms Levenberg-Marquardt (LM) algorithm [14] is performed. This approach provides a gradient based technique allowing fast error minimization. The major aim of training is to get the appropriate values of the weights for closest possible prediction through repetitive iterations. The LM method works on the principle of minimizing the mean squared error between actual output of the system and predicted output of the network and can be calculated with the following formula.

$$V_N(\theta) = \frac{1}{N} \sum_{t=1}^{N} (y(t) - \hat{y}(t))^2 \tag{1}$$

Where

$$\hat{y}(t) = g(\theta, \phi(t)) \tag{2}$$

$$\theta = (a_1, a_2, \ldots a_{na}, b_1, b_2, b_{nb}) \tag{3}$$

$$\phi(t) = (y(t-1), \ldots y(t-na), u(t-nk), \ldots u(t-nk-nb+1)) \tag{4}$$

Here $\phi$ is the matrix of past inputs and outputs. To find the coefficient $\theta$, there are many assumptions and detailed knowledge of the plant is required.

In each of the three control architectures mention above, system identification stage is identical but control design stage is different. For model predictive control, the plant model is used to predict future behavior of the plant, and an optimization algorithm is used to select the control input that optimizes future performance.

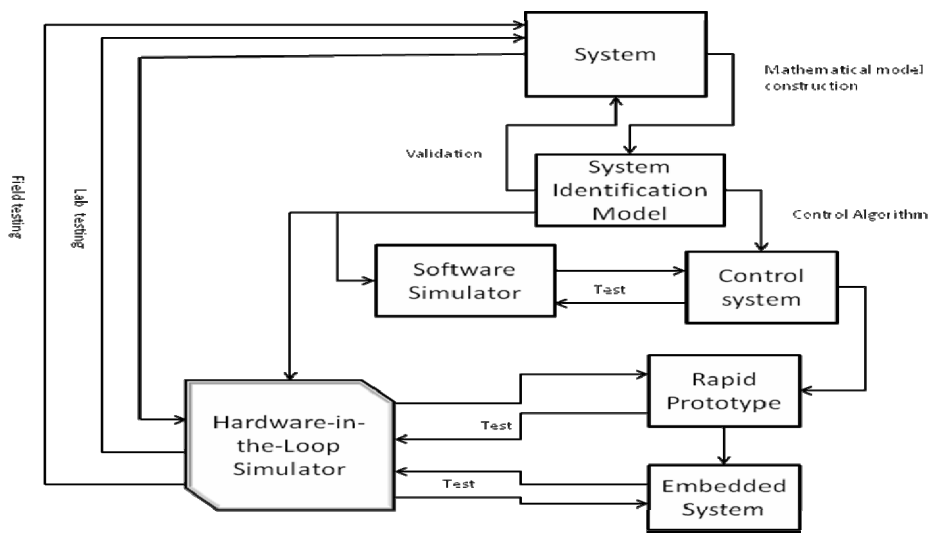

Fig. 5. Flow of Control System Design with System Identification

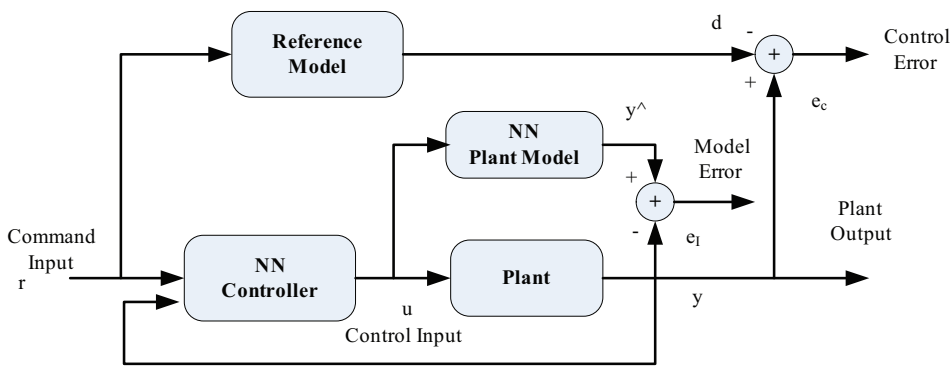

Fig. 6. Neural Network MRC architecture

For NARMA-L2 control, the controller is simply a rearrangement of the plant model. We used model reference control to simulate the nonlinear identification and control of UAV. For model reference control, the controller is a neural network that is trained to control a plant so that it follows a reference model. The neural network plant model is used to assist in the controller training.

The neural model reference control architecture uses two neural networks: a controller network and a plant model network, as shown in the Fig. 6. The plant model is identified first, and then the controller is trained so that the plant output follows the reference model output. The system identification error can be defined by

$$e_I = y - \hat{y} \qquad (5)$$

and the tracking error

$$e_c = y - d \qquad (6)$$

Controller parameters are updated based on error computed from the system output and the NN model of the plant.

They describe the input /output behavior of the system using a set of weights. Such models can be interpreted as a weighted combination of several local models resulting in a nonlinear global model. Hence the mismatch between the nonlinearities of local models and process is less significant compared with single nonlinear model. Therefore neural network modeling has been applied especially to modeling tasks with uncertain nonlinearities, uncertain parameters and or high complexity.

In the case of multi-input multi-output (MIMO) plants, the plant identification stage is same as that for single-input single-output (SISO) except that the NN model has many neurons in the output layer as the number of outputs of actual plant. Fig. 7 shows the control of a MIMO plant using NN controller.

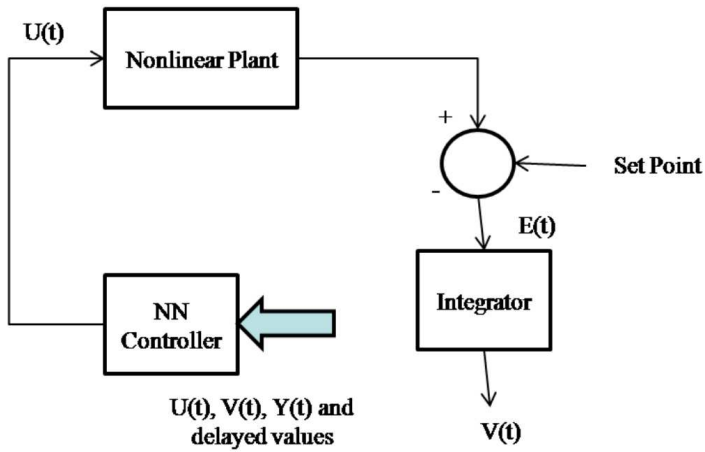

Fig. 7. Control of a MIMO plant using NN controller

### 3.1 Training of model using NN

Iterative training is conducted to minimize mean square error (MSE) using Levenberg Marquardt (LM) algorithm. The LM is gradient based approach that allows fast error minimization. The mission of training is to obtain the most suitable and optimized values of the weights for closest prediction through iterations.

The training process (Fig 8) is an iterative and can be stopped either when total training error reaches a bottom threshold or when training error ceases to decrease any further. There is flexibility for varying number of neurons in the hidden layers to optimize error. Starting from a small number of neurons, the number can be gradually increased or decrease until an accepted training error is achieved. Once the NN is successfully trained, it can be used to obtain a linear model of the plant from the available input and output values. Figure shows the behavior of plant input and output. It can be seen that the output response is different in each time slot with the variation of input weights. As a typical case, 1000 different sets of initial weights are considered for the network with 10 hidden neurons.

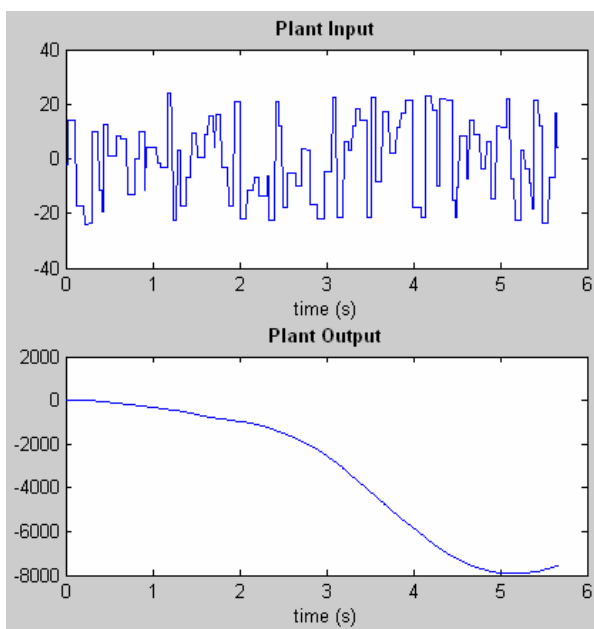

Fig. 8. Plant Input/output behavior during training

The network architecture for MIMO model of NN is shown in figure 9. In a particular equilibrium condition, no inertial and aerodynamic coupling, the behavior of RUAV can be divided into lateral and longitudinal dynamics mode and train with MIMO. The longitudinal cyclic deflection and collective control input is used to control the longitudinal dynamics mode whereas lateral cyclic deflection and pedal control input is used to control the lateral dynamics mode. The NN has trained with four system inputs and six outputs. The number of hidden layer is considered. The network is trained with different sets of data collected from the real flight tests of the RUAV.

Figure 9 describes the longitudinal dynamics mode as longitudinal cyclic deflection and collective control are provided as inputs (U) to the system and pitch rate and forward velocity (u) are considered as the outputs (Y) of the system that results four outputs pitch angle ($\theta$), forward velocity (u),vertical velocity (w) and pitch angular rate (q). Similarly, lateral cyclic deflection and pedal control are provided as inputs to the system and vertical velocity and roll rate are considered as the outputs of the system that results roll angle ($\varphi$), lateral velocity (v), roll angular rate (p), yaw angular rate (r).

Figure 10 illustrates a scenario of getting performance behavior of the identification with plant process during training (21 Epochs), where the training data and testing data are following almost same behavior.

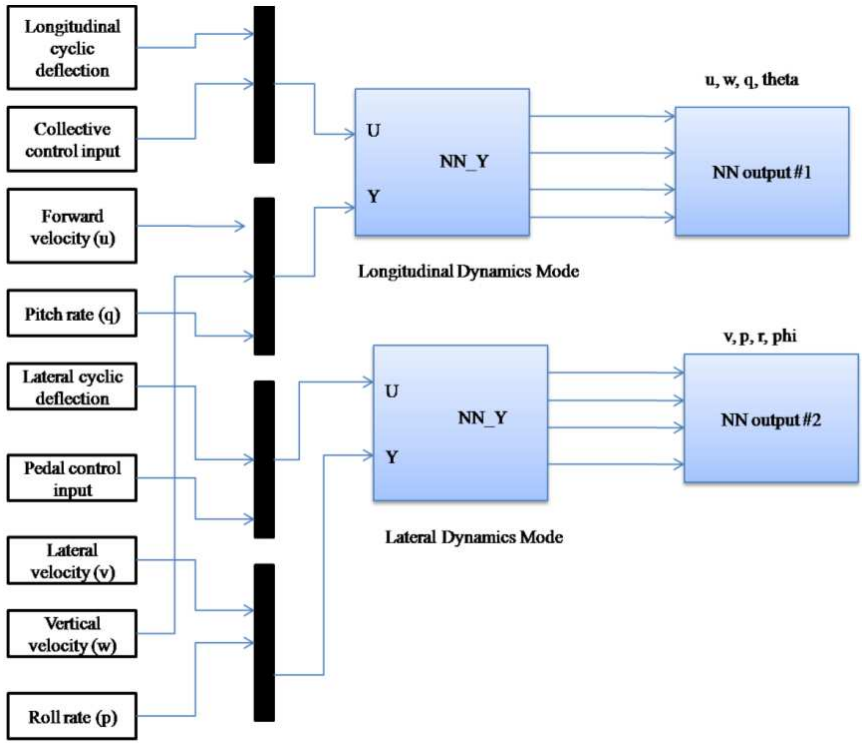

Fig. 9. MIMO model of NN

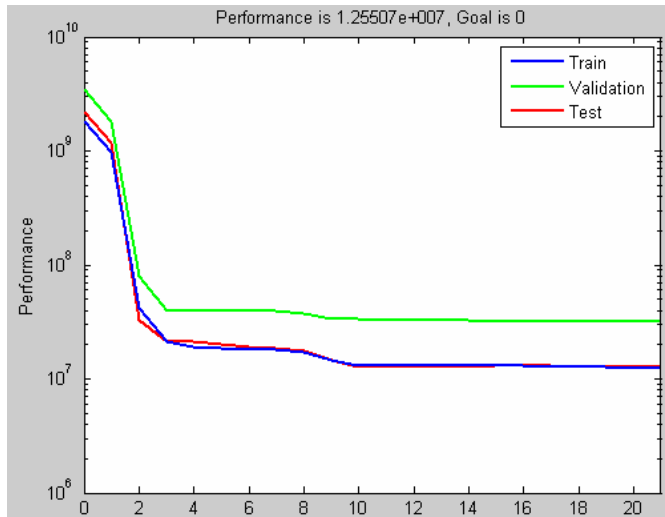

Fig. 10. A scenario of getting performance behavior of the identification with plant process during training (Trainlm at 21 Epochs), where the training data and testing data are following almost same behavior

## 4. Design of control system

The key challenge to deploy the designed control system of UAV is the potential risk and cost. So, to minimize the cost and the potential risk, we have to test and simulate the control system rigorously to get high degree of precision of safeness. Design, testing and simulation are the iterative process and this can be accomplished by executing several pair of design, test, and simulation. The simulation of UAV control system consist a group of nonlinear Simulink models which is used to estimate the capabilities of controllers. Generally, these models are employed both in evolutionary search to estimate the robustness of a particular controller, and later to verify and validate the designed controllers through extensive simulation with several test cases in different conditions. An UAV represents a complex non-linear system with 6 Degree of Freedom (6-DOF), and having high degree of coupling. It is anticipated that the most effective control on such a system can be gained with an appropriate non-linear controller.

NNs have attracted a great deal of attention owing to their ability to learn non-linear functions from input-output data examples [8]. Applied to control field, NNs are essentially nonlinear models that can be useful to solve non-linear control problems [9].

### 4.1 Mathematical model of RUAV dynamics

Basic starting point for UAV control design is to find out the state space matrix from 6-DOF equations of motion by linearizing with proper assumptions. The state of a system is a set of variables (Fig. 11) such that the knowledge of these variables and the input functions will, with the equations describing the dynamics, provide the future state and output of the system. The state of the system is described by the set of the first-order differential equations written in terms of state variables [x1 x2 ………xn].

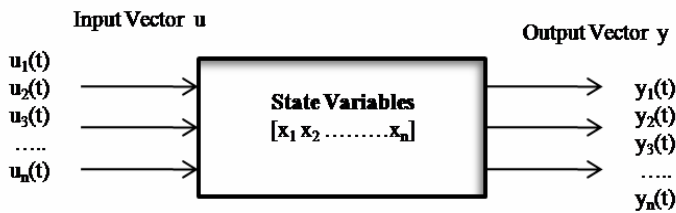

Fig. 11. System inputs and outputs

The state space is defined as the n-dimensional space in which the components of the n state vector represent its coordinate axes. The state equations of a system are a set of n first-order differential equations, where n is the number of independent states. Many control problems, however, that require multiple outputs be controlled simultaneously, to do control of such system, multiple inputs must be manipulated, usually they are orchestrated as MIMO. The helicopter is a complex MIMO system with high correlation. UAV autopilot is an example of MIMO where speed, altitude, pitch, roll, and yaw angles must be maintained and throttle, several rudders, and flaps are available as control variables.

The UAV systems consist of a six degree of freedom, nonlinear complex systems. Budiyono A. et al., [11] illustrate the nonlinear rigid body equations of motion of helicopter (Eq.7-15) that describes the vehicle's translational motion and angular motion about three reference axes.

$$\sum X = m(\vec{u} - rv + qw) + mg\sin\theta \tag{7}$$

$$\sum Y = m(ru + \vec{v} - pw) - mg\sin\phi\cos\theta \tag{8}$$

$$\sum Z = m(-qu + pv - \vec{w}) - mg\cos\phi\cos\theta \tag{9}$$

$$\sum L = I_{xx}\vec{p} - (I_{yy} - I_{zz})qr \tag{10}$$

$$\sum M = I_{yy}\vec{q} - (I_{zz} - I_{xx})pr \tag{11}$$

$$\sum N = I_{zz}\vec{r} - (I_{xx} - I_{yy})pq \tag{12}$$

$$\vec{\phi} = p + (q\sin\phi + r\cos\phi)\tan\theta \tag{13}$$

$$\vec{\theta} = q\cos\phi - r\sin\phi \tag{14}$$

$$\vec{\varphi} = (q\sin\phi + r\cos\phi)\sec\theta \tag{15}$$

Where the vector u, v, w and p, q, r are the fuselage velocities and angular rates in the body coordinate system, respectively. X, Y, Z are the external forces acting on the helicopter center of gravity and L, M, N are the external moments. State space and transfer-function models can be generalized to MIMO models. These first-order differential equations can be written in a general form that can be represented in matrix notation.

$$\vec{x} = A\vec{x} + B\vec{u}$$
$$\vec{y} = C\vec{x} + D\vec{u} \tag{16}$$

Where

$$\vec{x} = \begin{bmatrix} u & w & q & \theta & a_{1s} & v & p & r & \phi & b_{1s} \end{bmatrix}' \tag{17}$$

And

$$\vec{u} = \begin{bmatrix} \delta_{long} & \delta_{coll} & \delta_{lat} & \delta_{ped} \end{bmatrix} \tag{18}$$

The MIMO transfer-function matrix can be obtained from state space model by $G(s) = C(sI - A)^{-1}B + D$ where $A \in \mathbb{R}^{n*n}, B \in \mathbb{R}^{n*m}, C \in \mathbb{R}^{l*n}, D \in \mathbb{R}^{l*m}$. The descriptions of all parameters are shown in Table 1 and Table 2.

Where A, B and C are the representation of the system matrix, input matrix and output matrix respectively. A, B, C and D depends on the flight regime with nominal parameter values for hovering and cruising. Then, u is a vector of the inputs, x is the element state vector, and y is a vector containing outputs. It is easy to see that each linear state space system of Equation (16) can be expressed as a linear time invariant (LTI) transfer functions. The procedure is to take Laplace transformation of the both sides of Equation (16) and use an algorithm is given by Leverrier-Fadeeva-Frame formula [10]. Another approach is to use Matlab functions directly. Let us take a transfer function of UAV (Eq. 19-20) to model and simulate.

| Parameter | Symbol | Description |
|---|---|---|
| Fuselage Linear Motion | u | Forward velocity |
| | v | Lateral velocity |
| | w | Vertical velocity |
| Fuselage Angular Motion | p | Roll Angular Rate |
| | q | Pitch Angular Rate |
| | r | Yaw Angular Rate |
| Rotor Tip-Path-Plane | a1s | Longitudinal Flapping Angle |
| | b1s | Lateral Flapping Angle |
| Pitch | $\theta$ | Pitch Angle |
| Roll | $\varphi$ | Roll Angle |

Table 1. Model states.

| Control | Description | Units |
|---|---|---|
| $\delta_{long}$ | Longitudinal Cyclic Deflection | Dimensionless [-1, 1] |
| $\delta_{lat}$ | Lateral Cyclic Deflection | Dimensionless [-1, 1] |
| $\delta_{ped}$ | Pedal control Input | Dimensionless [-1, 1] |
| $\delta_{coll}$ | Collective Control Input | Dimensionless [0, 1] |

Table 2. Control input variables.

$$A = \left[\begin{array}{ccccc|ccccc}
X_u & 0 & 0 & -g & -g & 0 & 0 & 0 & 0 & 0 \\
0 & Z_w & 0 & 0 & Z_{a1s} & 0 & 0 & Z_r & 0 & Z_{b1s} \\
M_u & 0 & 0 & 0 & M_{a1s} & M_v & 0 & 0 & 0 & M_{b1s} \\
0 & 0 & 1 & 0 & 0 & 0 & 0 & 0 & 0 & 0 \\
0 & 0 & -1 & 0 & A_{a1s} & 0 & 0 & 0 & 0 & A_{b1s} \\
\hline
0 & 0 & 0 & 0 & 0 & Y_v & 0 & 0 & g & g \\
L_u & 0 & 0 & 0 & L_{a1s} & L_v & 0 & 0 & 0 & L_{b1s} \\
0 & N_w & 0 & 0 & 0 & 0 & N_p & N_r & 0 & 0 \\
0 & 0 & 0 & 0 & 0 & 0 & 1 & 0 & 0 & 0 \\
0 & 0 & 0 & 0 & B_{a1s} & 0 & -1 & 0 & 0 & B_{b1s}
\end{array}\right] \tag{19}$$

$$B = \left[\begin{array}{cccc}
0 & 0 & 0 & 0 \\
Z_{col} & 0 & 0 & 0 \\
M_{col} & 0 & 0 & 0 \\
0 & 0 & 0 & 0 \\
0 & A_{lon} & 0 & A_{lat} \\
0 & 0 & Y_{ped} & 0 \\
0 & 0 & 0 & 0 \\
N_{col} & 0 & N_{ped} & 0 \\
0 & 0 & 0 & 0 \\
0 & B_{lon} & 0 & B_{lat}
\end{array}\right] \tag{20}$$

$$C = \begin{bmatrix} 1 & 0 & 0 & 0 & 0 & 0 & 0 & 0 & 0 & 0 \\ 0 & 1 & 0 & 0 & 0 & 0 & 0 & 0 & 0 & 0 \\ 0 & 0 & 1 & 0 & 0 & 0 & 0 & 0 & 0 & 0 \\ 0 & 0 & 0 & 1 & 0 & 0 & 0 & 0 & 0 & 0 \\ 0 & 0 & 0 & 0 & 1 & 0 & 0 & 0 & 0 & 0 \\ 0 & 0 & 0 & 0 & 0 & 1 & 0 & 0 & 0 & 0 \\ 0 & 0 & 0 & 0 & 0 & 0 & 1 & 0 & 0 & 0 \\ 0 & 0 & 0 & 0 & 0 & 0 & 0 & 1 & 0 & 0 \\ 0 & 0 & 0 & 0 & 0 & 0 & 0 & 0 & 1 & 0 \\ 0 & 0 & 0 & 0 & 0 & 0 & 0 & 0 & 0 & 1 \end{bmatrix} \tag{21}$$

$$D = \begin{bmatrix} 0 & 0 & 0 & 0 \\ 0 & 0 & 0 & 0 \\ 0 & 0 & 0 & 0 \\ 0 & 0 & 0 & 0 \\ 0 & 0 & 0 & 0 \\ 0 & 0 & 0 & 0 \\ 0 & 0 & 0 & 0 \\ 0 & 0 & 0 & 0 \\ 0 & 0 & 0 & 0 \\ 0 & 0 & 0 & 0 \end{bmatrix} \tag{22}$$

The descriptions of all parameters used in Eq. 19-20 are shown in Table 3

| Parameter | Description |
|---|---|
| Zw Zr Zbls Zals Yv Xu Nw Nr Np Mv Mu Mbls Mals Lv Lu Lals Lbls Bals Bbls Aals Abls | Stability derivative |
| g | Force of gravity |
| Zcol Yped Ncol Nped Mcol | Control derivative |
| Blon Blat Alat Alon | Cyclic input sensitivity |

Table 3. Parameters of model constants for fuselage linear motion equations, model constants for tip-path-plane and augmented yaw dynamics, and model constants for angular motion.

The objective of training a NN is to minimize the error between the output of NN and the desired output. First, we use models (Eq. 19-20) to generate training data. Then by propagation algorithm, all the weights in NN plant can be adjusted through the training sets until the NN plant outputs are very close to the plant outputs. This completes the system identification. Second, we will choose a reference model which allows the desired behavior. Let us use a flight data from Eq. 23-24 for design and simulation.

$A =$

$$\begin{bmatrix} -0.78501 & 0 & 0 & -9.8 & -9.8 & 0 & 0 & 0 & 0 & 0 \\ 0 & -0.065145 & 0 & 0 & -56.659 & 0 & 0 & -0.79784 & -0.0045036 & 1344.1 \\ 0.35712 & 0 & 0 & 0 & 92.468 & -0.063629 & 0 & 0 & 0 & 56.515 \\ 0 & 0 & 1 & 0 & 0 & 0 & 0 & 0 & 0 & 0 \\ 0 & 0 & -1 & 0 & -11.842 & 0 & 0 & 0 & 0 & -7.1176 \\ 0 & 0 & 0 & 0 & 0 & 0.11245 & 0 & 0 & 9.8 & 9.8 \\ 0.46624 & 0 & 0 & 0 & -0.6588 & -0.083441 & 0 & 0 & 0 & 131.19 \\ 0 & 1.0349 & 0 & 0 & 0 & 0 & -9.9435 & -0.30115 & 0 & 0 \\ 0 & 0 & 0 & 0 & 0 & 0 & 1 & 0 & 0 & 0 \\ 0 & 0 & 0 & 0 & 2.1755 & 0 & -1 & 0 & 0 & -14.687 \end{bmatrix} \quad (23)$$

$B =$

$$\begin{bmatrix} 0 & 0 & 0 & 0 \\ 0.71986 & 0 & 0 & 0 \\ 1.4468 & 0 & 0 & 0 \\ 0 & 0 & 0 & 0 \\ 0 & -11.198 & 0 & 4.3523 \\ 0 & 0 & 204.28 & 0 \\ 0 & 0 & 0 & 0 \\ -3.5204 & 0 & -7.5159 & 0 \\ 0 & 0 & 0 & 0 \\ 0 & 2.9241 & 0 & 11.712 \end{bmatrix} \quad (24)$$

**Eigenvalues:**

The key dynamics can be seen from the system's Eigen values and Eigen vectors, are listed in Table 4. The system is stable with damping because all the real parts of the eigenvalues are negative. The simulation (Fig. 12) shows clearly that the system is stable but having damping.

| Sno | Eigenvalue | Damping | Freq. (rad/s) |
|-----|------------|---------|---------------|
| 1 | -2.01e-002 + 8.27e-003i | 9.25e-001 | 2.17e-002 |
| 2 | -2.01e-002 - 8.27e-003i | 9.25e-001 | 2.17e-002 |
| 3 | -1.83e-001 + 9.01e-001i | 1.99e-001 | 9.19e-001 |
| 4 | -1.83e-001 - 9.01e-001i | 1.99e-001 | 9.19e-001 |
| 5 | -2.82e-001 + 5.79e-001i | 4.37e-001 | 6.44e-001 |
| 6 | -2.82e-001 - 5.79e-001i | 4.37e-001 | 6.44e-001 |
| 7 | -5.93e+000 + 6.22e+000i | 6.90e-001 | 8.59e+000 |
| 8 | -5.93e+000 - 6.22e+000i | 6.90e-001 | 8.59e+000 |
| 9 | -7.37e+000 + 1.06e+001i | 5.73e-001 | 1.29e+001 |
| 10 | -7.37e+000 - 1.06e+001i | 5.73e-001 | 1.29e+001 |

Table 4. Eignevalues of the helicopter system

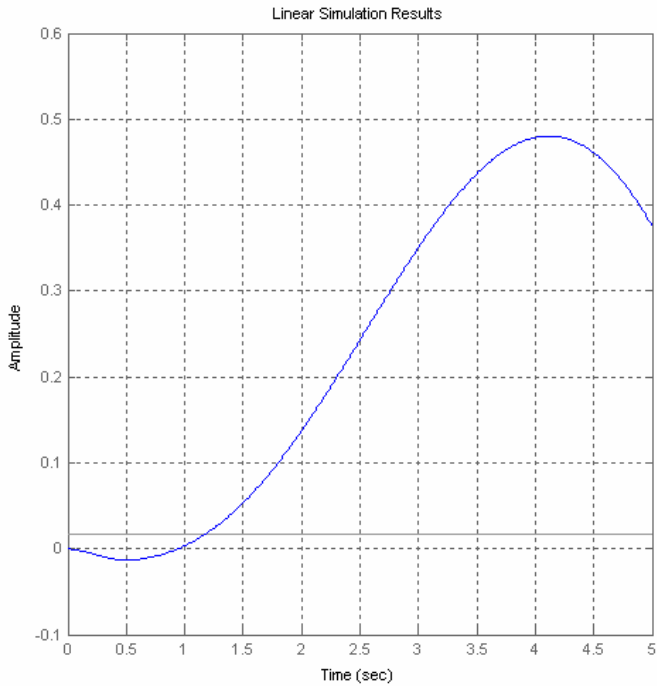

Fig. 12. Transfer function of system

## 5. Experimental results and analysis

In this experiment we used NN approach to train MIMO model and capture the phenomena of flight dynamics. This simulation is divided into two parts longitudinal mode and Lateral mode. The NN approach considers separate lateral and longitudinal network with inertial coupling between the networks taken into consideration. These networks trained individually by making it MIMO model. Basically system identification process consists of gathering experimental data, estimate model from data and validate model with independent data. NN controller is designed in such a way that makes the plant output to follow the output of a reference model. The main target is to play with fine tuning of controller in order to minimize the state error.

The experiment is carried out with System identification procedures with Prediction Error Method (PEM) algorithm using System Identification Toolbox using Levenberg-Marquardt (LM) algorithm. We observe NN approach to get better result of System identification that shows the perfect matching and shown as RUAV Longitudinal Dynamics and RUAV Lateral Dynamics in the following fig. 13-18

The prediction error of the output responses is described in Fig. 14. The autocorrelation function almost tend to zero and the cross correlation function vary in the range of -0.1to 0.1. This shows the dependency between prediction error and $\delta_{coll}$, $\delta_{long}$ but the dependency rate is very less.

**Longitudinal Dynamics Mode Analysis**

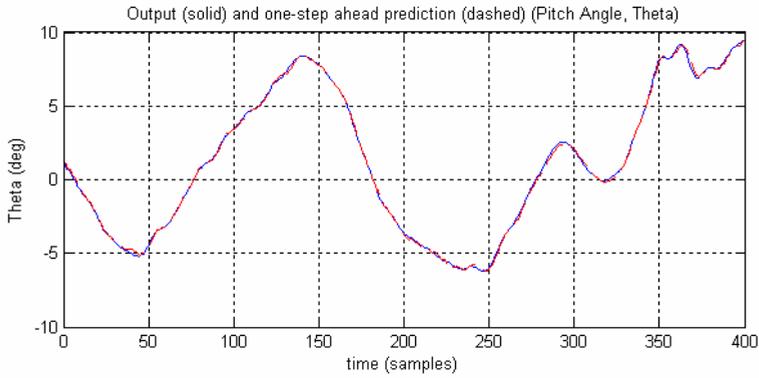

(a)  Pitch Angle ($\theta$)

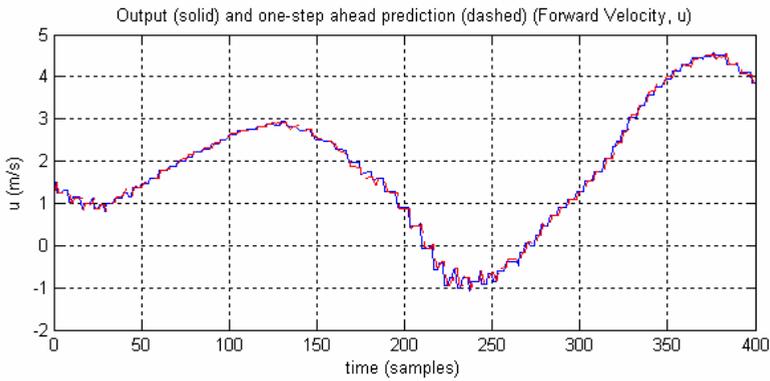

(b) Forward Velocity (u)

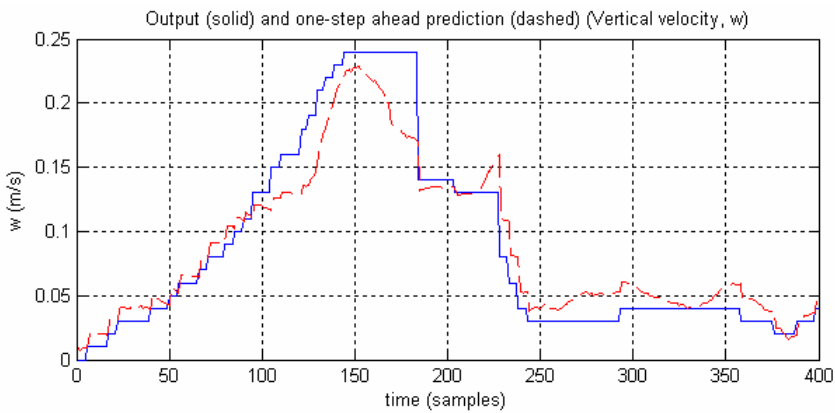

(c) Vertical velocity (w)

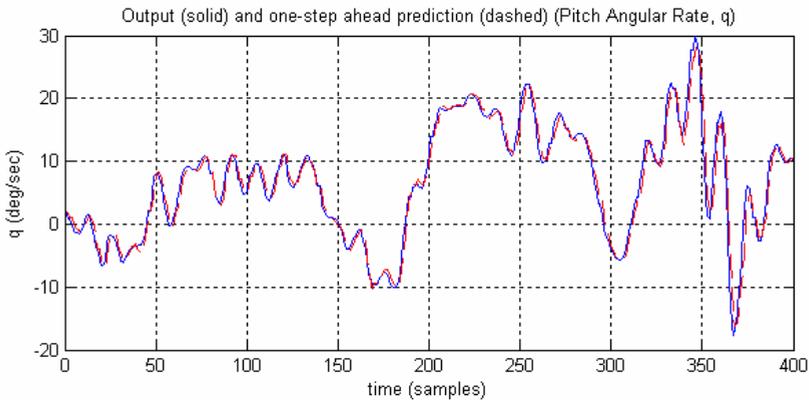

(d) Pitch Angular Rate (q)

Fig. 13. Output response with network response in Longitudinal dynamics mode

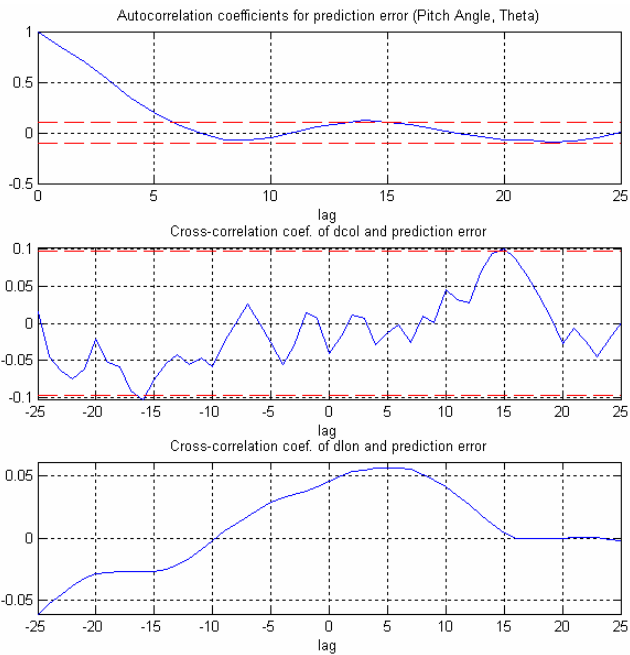

(a) Pitch Angle ($\theta$)

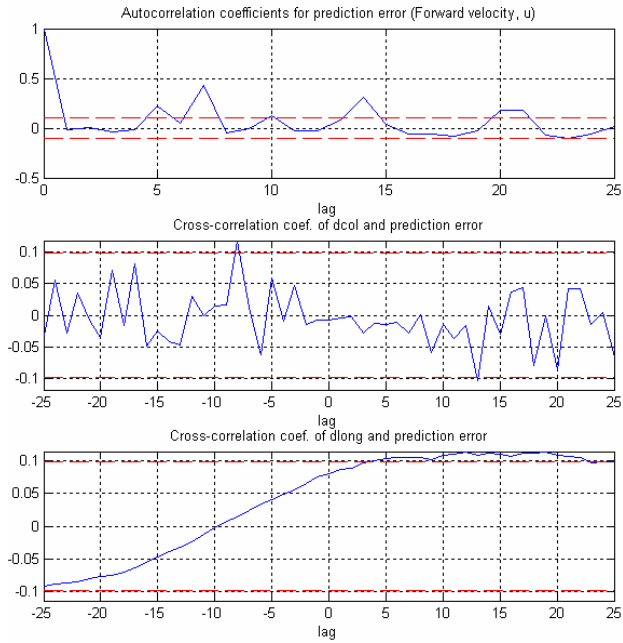

(b) Forward velocity (u)

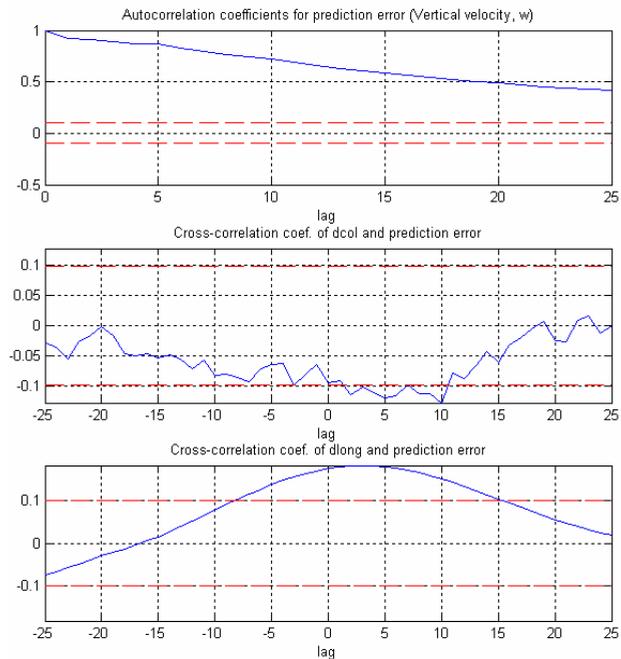

(c) Vertical velocity (w)

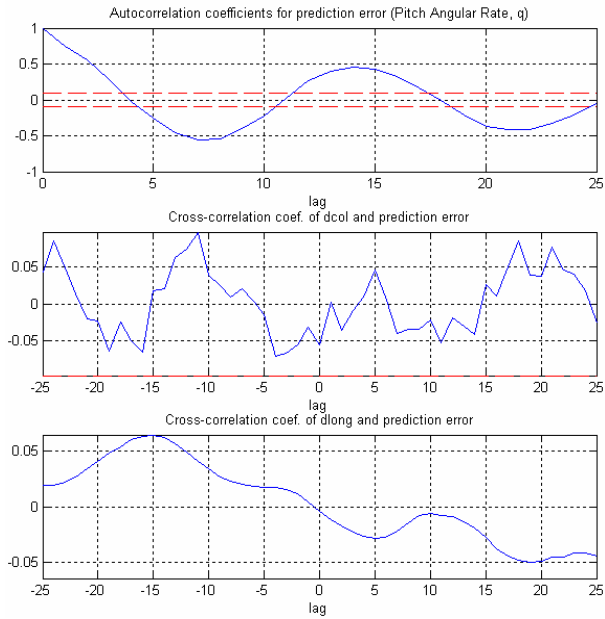

(d) Pitch Angular Rate (q)

Fig. 14. Autocorrelation and Cross-correlation of output response in longitudinal mode
The histogram of prediction error is shown in Fig. 15.

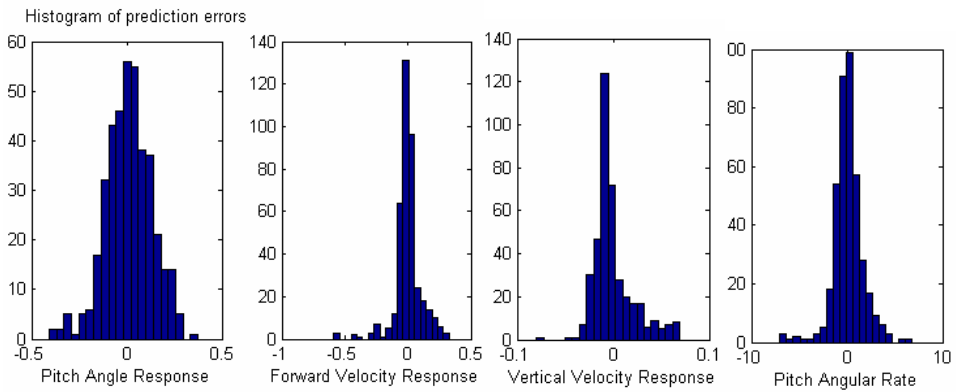

Fig. 15. Histogram of Prediction errors in Longitudinal Mode

**Lateral Dynamics Mode Analysis**

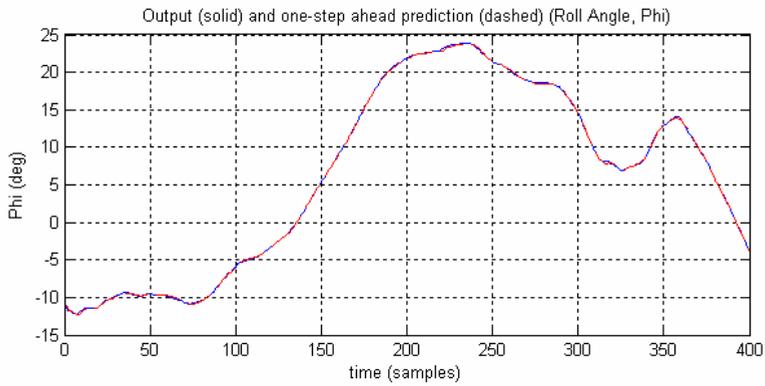

(a) Roll Angle ($\varphi$)

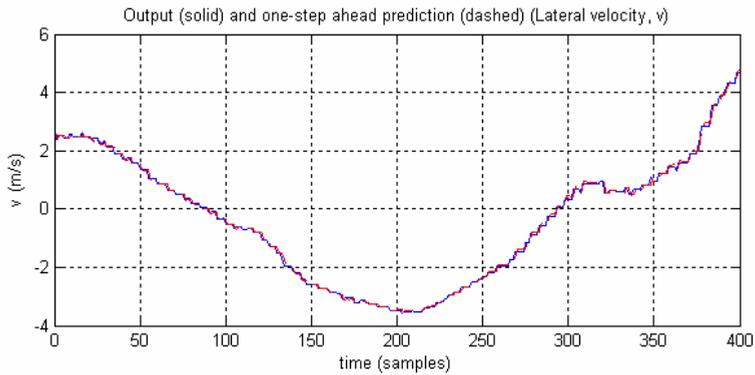

(b) Lateral Velocity (v)

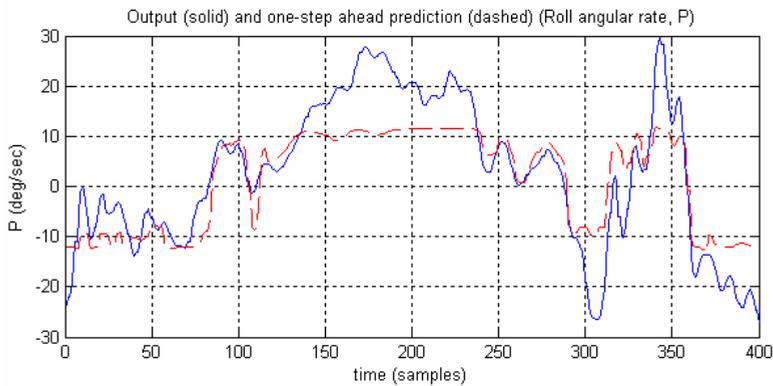

(c) Roll Angular Rate (P)

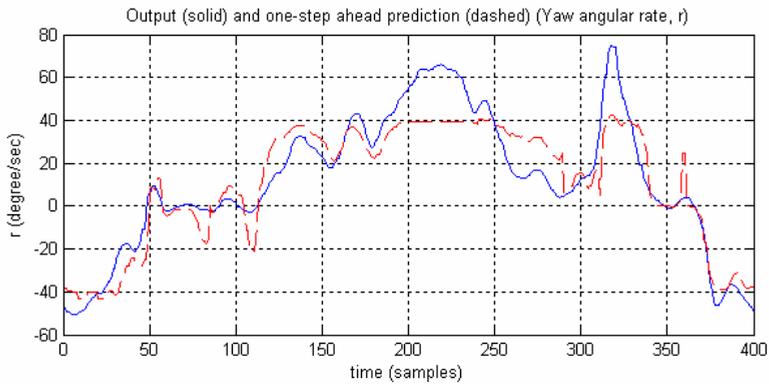

(d) Yaw Angular Rate (r)

Fig. 16. Output response with network response in lateral dynamics mode

The prediction error of the output responses is described in Fig. 17. Similarly, in lateral mode also, the autocorrelation function almost tend to zero and the cross correlation function vary in the range of -0.1 to 0.1. This shows the dependency between prediction error and $\delta_{lat}$, $\delta_{ped}$ but the dependency rate is very less.

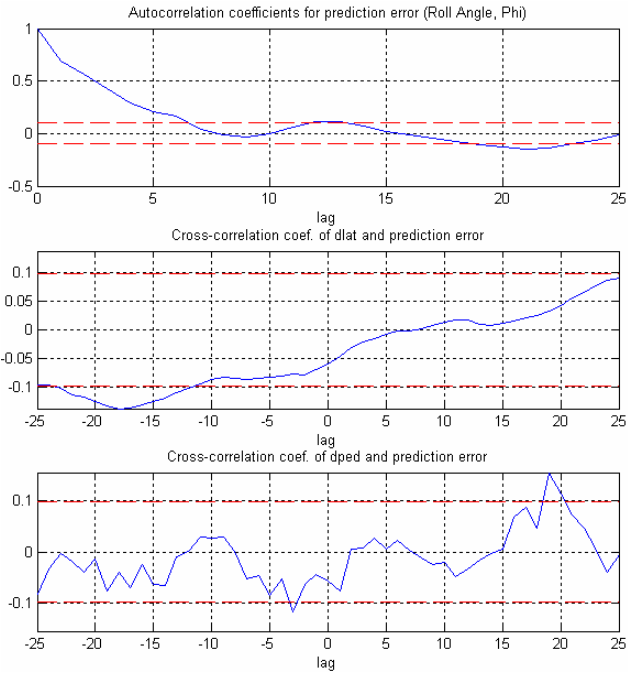

(a) Roll Angle ($\varphi$)

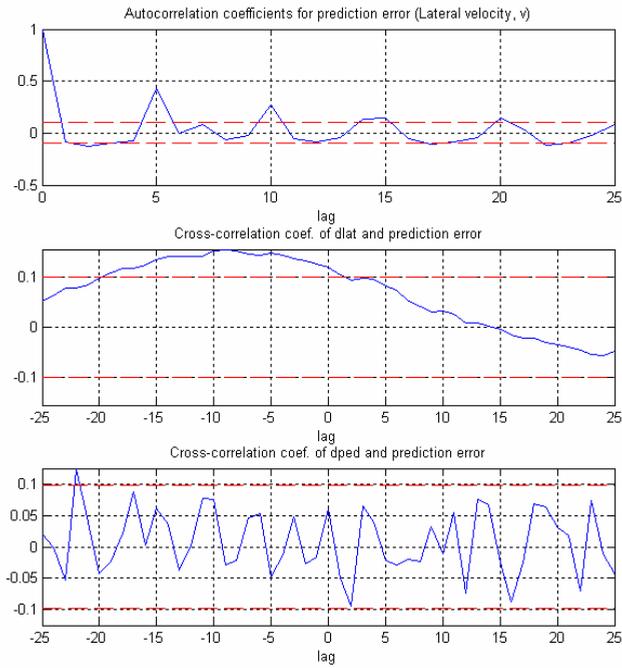

(b) Lateral Velocity (v)

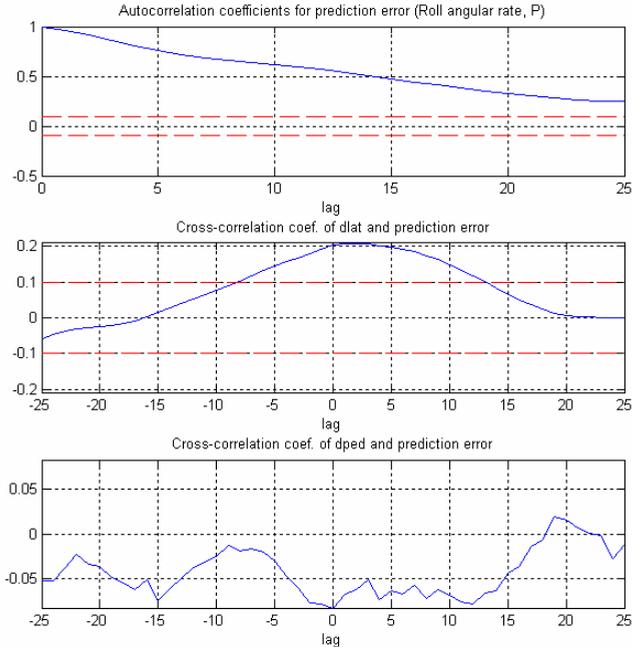

(c) Roll Angular Rate (P)

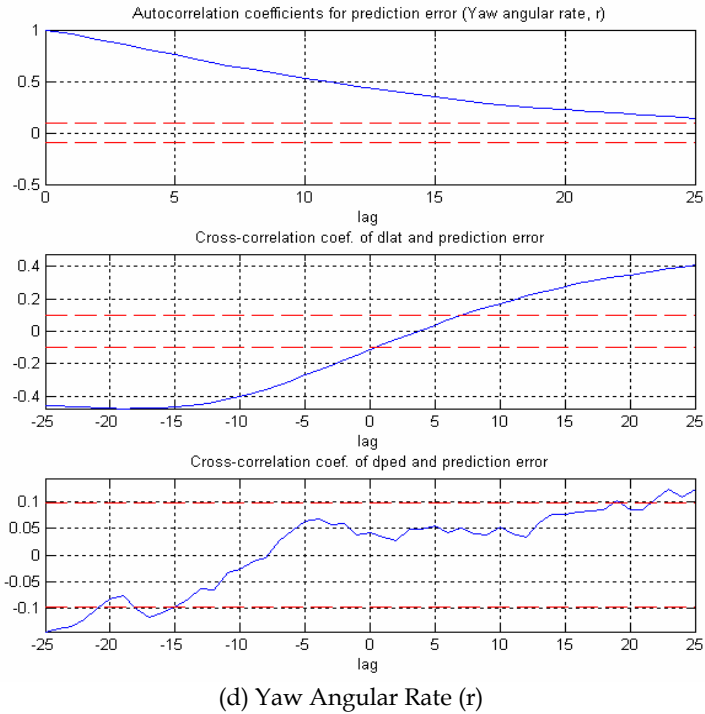

(d) Yaw Angular Rate (r)

Fig. 17. Autocorrelation and Cross-correlation of output response in lateral mode

The histogram of prediction error is shown in Fig. 18.

## 6. Conclusion

UAV control system is a huge and complex system, and to design and test a UAV control system is time-cost and money-cost. This chapter considered the simulation of identification of a nonlinear system dynamics using artificial neural networks approach. This experiment develops a neural network model of the plant that we want to control. In the control design stage, experiment uses the neural network plant model to design (or train) the controller. We used Matlab to train the network and simulate the behavior.

This chapter provides the mathematical overview of MRC technique and neural network architecture to simulate nonlinear identification of UAV systems. MRC provides a direct and effective method to control a complex system without an equation-driven model. NN approach provides a good framework to implement MEC by identifying complicated models and training a controller for it.

## 7. Acknowledgment

"This research was supported by the MKE (Ministry of Knowledge and Economy), Korea, under the ITRC (Information Technology Research Center) support program supervised by the NIPA (National IT Industry Promotion Agency)" (NIPA-2010-C1090-1031-00003)

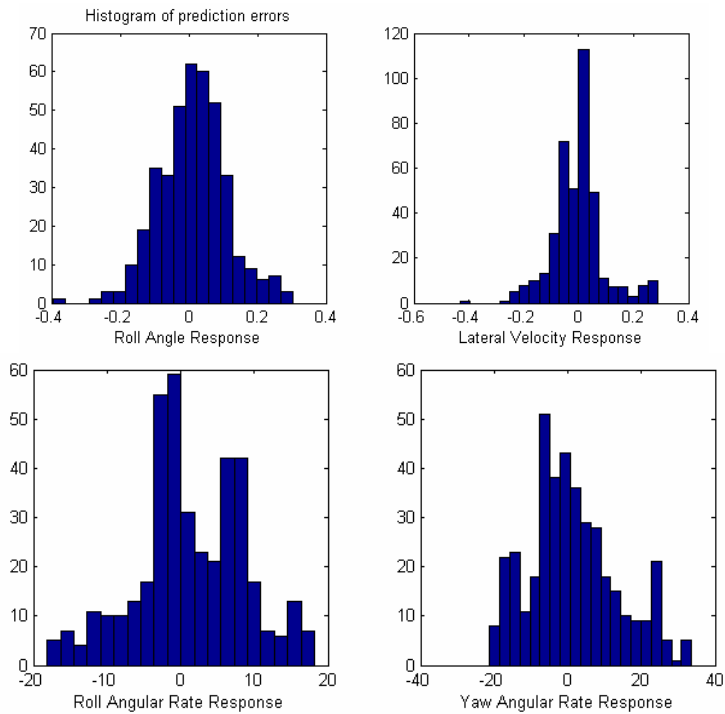

Fig. 18. Histogram of Prediction errors in Longitudinal Mode